\documentclass[11pt,a4paper]{iopart}
\usepackage[utf8]{inputenc}
\usepackage{url}
\usepackage{graphicx}
\usepackage{iopams}
\usepackage{setstack}
\usepackage{color}
\usepackage{bm}
\usepackage{arydshln}
\usepackage{enumitem}

\makeatletter
\long\def\dddddot#1{%
  {\mathop {#1}\limits ^{\vbox to-1.4\ex@ {\kern -\tw@ \ex@ \hbox {\normalfont .....}\vss }}}%
}
\long\def\multidots#1#2{%
  \count@=0
  {{\mathop {#2}\limits ^{\vbox to-1.4\ex@ {\kern -\tw@ \ex@ \hbox {\normalfont %
  \loop%
  \ifnum#1>\count@%
  .%
  \advance\count@ by1%
  \repeat%
  }\vss }}}}%
}
\makeatother

\def\d{{\rm d}}

\begin{document}

\title{Quantum signatures from Ho\v rava-Lifshitz cosmography}

\author{Celia Escamilla-Rivera}
\address{Instituto de Ciencias Nucleares, Universidad Nacional Aut\'{o}noma de M\'{e}xico, 
Circuito Exterior C.U., A.P. 70-543, M\'exico D.F. 04510, M\'{e}xico.}
\ead{celia.escamilla@nucleares.unam.mx}

\author{Hernando Quevedo}
\address{Instituto de Ciencias Nucleares, Universidad Nacional Aut\'onoma de M\'exico, Circuito Exterior C.U., A.P. 70-543, M\'exico D.F. 04510, M\'exico.}
\address{Dipartimento di Fisica and ICRA,  Universit\`a di Roma ``La Sapienza", I-00185 Roma, Italy}
\address{Institute for Experimental and Theoretical Physics,  Al-Farabi Kazakh National University,  Almaty 050040, Kazakhstan}
\ead{quevedo@nucleares.unam.mx}

\date{\today}

\begin{abstract}
In this paper we propose an alternative cosmography by considering Ho\v rava-Lifshitz gravity as a model of quantum gravity to search for quantum effects at the cosmological level. For our analyses we consider current late universe surveys and a Gravitational Waves forecast from Einstein Telescope. We found naturally a non-flat scenario with $\Omega_k =-0.021^{+0.023}_{-0.029}$ with $H_0 = 71.904^{+1.406}_{-1.347}$, without showing the standard reported $3.4$-$\sigma$ inconsistency. Furthermore, we obtained a specific value for the Ho\v rava parameter 
$\omega \approx -3.8\times 10^{-14}$, which can be interpreted as a measure of a quantum effect and could be used to further test this quantum gravity model. We conclude that classically, according to the $\Lambda$CDM model, our universe seems to be spatially flat, but in fact, it is curved from a quantum point of view.
\end{abstract}

\maketitle


\section{Introduction}

Obtaining a theory of Quantum Gravity (QG) is still an unsolved problem in modern physics \cite{carlip01,kiefer05}. The objective one has in mind when making this statement is obtaining a theory of gravity which is valid at all scales. Since gravity is a rather weak force, its quantum effects are not expected to be observable at scales which we can currently probe. The criteria by which a candidate theory for QG may be assessed are limited to mathematical/internal consistency, reproducing the classical descriptions of General Relativity (GR) and leading to nontrivial predictions that can eventually be tested.
As it is standard, we try to implement quantisation methods to GR. However, GR is perturbatively non-renormalizable and new methods have to be found. 
One option is to follow a minimal approach and obtain a perturbatively renormalizable UV completion of GR as a Quantum Field Theory (QFT) for the metric field in 3+1 dimensions. Although, this proposal requires giving up some properties of GR, as its original formulation does not satisfy the perturbative renormalisability requirement, at least not at the Gaussian fixed point. 
This line of thought can be viewed as the motivation for a theory of QG with anisotropic scaling in the UV, the so-called Ho\v rava-Lifshitz gravity (HLG) \cite{hor09a,hor09b}, where the treatment of the UV fixed point is akin to what is done in critical systems.

In a recent work \cite{Luongo:2018oil}, HLG was tested in the context of cosmography. Essentially, cosmography assumes only the validity of the Cosmological Principle, which fixes the geometry of spacetime, without invoking any particular dynamical theory. Therefore, it is a powerful tool to test the theoretical predictions of dynamical models and to compare different models concerning observations. In this analysis, HLG turns out to be disfavored when compared with the standard cosmological model, which follows from Einstein gravity. 
For our proposal, we perform an alternative analysis by considering HLG as 
the classical limit of a not yet formulated  model of quantum gravity and search for quantum signatures at the cosmological level. Certainly, quantum effects are expected to be dominant in the very early universe and very small in the present time. Nevertheless, we use in this work current time observations to search for detectable effects in a way that resembles the analysis of the cosmic microwave background radiation as the present signature of the initial Big Bang singularity. As long as no realistic quantum model for the origin of the Universe is known, we must use classical models such as HLG, from which we expect to contain footprints of quantum effects due to its property of being renormalizable. 

\color{black}
 To investigate cosmology in HLG,  we use the new formalism called \textit{inverse cosmography} \cite{Munoz:2020gok} with recent and simulated observational data, including a higher redshift range, from which we do not exclude {\it a priori} the possibility of having quantum prints. Then, we analyse HLG to establish signatures that could follow from the quantum nature of the model. In these lines of thought, the inverse cosmography was proposed to test the cosmodynamical parameters and use them to analyse the kinematics via its generic cosmographic parameter, this without experimenting with the standard problems of truncation of the cosmographic series. 

\section{Standard cosmography description} 

The cosmographic method can be described as an independent technique to obtain viable limits on the cosmic accelerated universe history at late times, considering by default the validation of the Cosmological Principle \cite{wei72,visser05}. Its requirements are
homogeneity and isotropy with spatial curvature given by a Friedmann-Lema{\^i}tre -Robertson-Walker (FLRW) metric
 $\d s^2=\d t^2-a(t)^2\left(\d r^2+r^2\d\Omega^2\right),$
where we employ $\d\Omega^2\equiv \d\theta^2+\sin^2\theta \d\phi^2$.
Usually, the cosmological scenario provides a whole energy density dominated by a constant $\Lambda$,
or by some dark energy density, with cold dark matter in second place and baryons as a small fraction on the rest. The corresponding
spatial curvature in the case of time-independent dark energy density is actually constrained to be negligible. 
This methodology can give us numerical results which do not
depend on the particular choice of the cosmological model, since only Taylor expansions are compared with observational samplers.
The cosmographic expansions can be obtained by determining the scale factor $a(t)$ as a Taylor series
around present time $t_0(z=0)$: 
\begin{eqnarray}
a(t)  & \sim & a(t_0)+ a'(t_0) \Delta t + \frac{a''(t_0)}{2} \Delta t^2+
\frac{a'''(t_0)}{6} \Delta t^3 
+   \frac{a^{(iv)}(t_0)}{24} \Delta t^4 +\frac{a^{(v)}(t_0)}{120} \Delta t^5+\ldots\,,
\end{eqnarray}
which recovers signal causality if one assumes $\Delta t\equiv t-t_0>0$. Therefore, from the latter expansion of $a(t)$, we can define
\begin{eqnarray}
H &\equiv& \frac{1}{a} \frac{\d a}{\d t}\,,\quad
q \equiv -\frac{1}{a H^2} \frac{\d^2a}{\d t^2}\,,\quad
j  \equiv \frac{1}{a H^3} \frac{\d^3a}{\d t^3}\,,
s \equiv \frac{1}{a H^4} \frac{\d^4a}{\d t^4}\,, \quad
l  \equiv \frac{1}{a H^5} \frac{\d^5a}{\d t^5}\,.
\end{eqnarray}

These equations are model-independent, i.e., they do not depend on the form of the dark energy fluid present in Friedmann equations
since they can be directly constrained by observations. They are the so-called: Hubble rate ($H$), acceleration parameter ($q$),  jerk parameter ($j$),  snap parameter ($s$), and lerk parameter ($l$), respectively.
Rewriting $a(t)$ in terms of these cosmographic parameters with a normalised scale factor to $a(t_0) = 1$
\begin{eqnarray}
a(t)  & \sim & 1+  H_0 \Delta t - \frac{q_0}{2}  H_0^2\Delta t^2+
+\frac{j_0}{6} H_0^3 \Delta t^3
+   \frac{s_0}{24}  H_0^4\Delta t^4  
+\frac{l_0}{120}  H_0^5\Delta t^5 +\ldots\,
\end{eqnarray}
We can re-scale the above cosmographic parameters in terms of the Hubble rate and its derivatives so that only three independent cosmographic coefficients appear explicitly
\begin{eqnarray}
\dot{H} &=& -H^2 (1 + q)\,,\\
\ddot{H} &=& H^3 (j + 3q + 2)\,,\\
\dddot{H} &=& H^4 \left [ s - 4j - 3q (q + 4) - 6 \right]\,,\\
\ddddot{H} &=& H^5 \left [ l - 5s + 10 (q + 2) j + 30 (q + 2) q + 24\right ]. \quad
\end{eqnarray}
Notice the correspondence degeneracy between the Hubble parameter derivatives and the cosmographic parameters
as a consequence that all these expressions are multiplied by $ H$.


\section{Background for inverse cosmography} 

In \cite{Munoz:2020gok}, it was pointed out that the degeneracy problem in cosmological models can be modulated using an \textit{inverse cosmography} methodology, which lies on the idea that the kinematics of the universe can be understood by setting limits of the current standard cosmography using not only Taylor-series like parameterisations, but also high order polynomials. This proposal also solves the problem of the error propagation over the statistical test, making it possible to set a cutoff on the cosmographic parameters directly from the dynamics of the cosmological ones. Therefore, we are directly seeing the dynamics of a model through its cosmographic parameters. 

We can set the equations to be used in the case of a specific model or, in our case, a model that can relax the UV limit. We derive the inverse cosmography considering a spatial flatness hypothesis on the Hubble function as
\begin{equation}\label{eq:friedmann_GR}
\left(\frac{H(z)}{H_0}\right)^2 =\Omega_k (1+z)^2 + \Omega_{m}(1+z)^3 + \Omega_r (1+z)^4 + \Omega_{i} f_{i}(z),
\end{equation}
where the curvature and radiation terms are $\Omega_k$ and $\Omega_r$, respectively. The term $\Omega_i$ is related with the standard description of the current universe evolution once we know the form of $f(z)$. Moreover, this approach requires a fiducial model, then we can write
a \textit{generic} expression for the cosmological equation of state (EoS)  \cite{Escamilla-Rivera:2019aol,Munoz:2020gok}
 \begin{equation}\label{eq:genericEoS}
w(z)= -1+\frac{1}{3} (1+z)\frac{f_{i}(z)^{\prime}}{f_{i}(z)}, 
\end{equation}
where the prime denotes $d/dz$. We note that the \textit{generic EoS} depends only on the form of $f_{i}(z)$. Following the inverse cosmography idea, we can use (\ref{eq:genericEoS}) with the chain rule
$\dot{}=d/dt =-(1+z)H(z)d/dz$, to obtain the set of cosmographic parameters in terms of $H(z)$ and its derivatives:
\begin{eqnarray}
q(z)= -1 +\frac{1}{2}(1+z)\frac{[H(z)^2]'}{H(z)^2}, \label{eq:q} 
\end{eqnarray}
\begin{eqnarray}
j(z) &=&\frac{1}{2}(1+z)^2 \frac{[H(z)^2]''}{H(z)^2}
 -(1+z)\frac{[H(z)^2]'}{H(z)^2} +1, \label{eq:j}\quad
\end{eqnarray}
\begin{eqnarray}
s(z) &=& -\frac{1}{6} (1+z)^3 \frac{[H(z)^2]'''}{H(z)^2} +\frac{1}{2}(1+z)^2 \frac{[H(z)^2]''}{H(z)^2} 
+(1+z)\frac{[H(z)^2]'}{H(z)^2} -1. \label{eq:s}
\end{eqnarray}
By solving and evaluating these expressions at $z=0$, we get the usual cosmographic series 
\begin{eqnarray}\label{eq:Hcosmo}
H(z)&=& H_0 +\frac{dH}{dz}\bigg |_{z=0} z +\frac{1}{2!}\frac{d^2 H}{dz^2}\bigg |_{z=0} z^2 
+\frac{1}{3!} \frac{d^3 H}{dz^3}\bigg |_{z=0} z^3 +\ldots %
\end{eqnarray}
Similar expressions can be calculated by expressing everything in terms of the function normalised by the Hubble constant, $E(z)=H(z)/H_0$ and its derivatives. Nonetheless, the information we can obtain is exactly equivalent. 


\section{Ho\v rava-Lifshitz dynamics} 

In HLG, it is  assumed that Lorentz invariance is violated at the UV regime and that the theory reduces to GR with  cosmological constant in the IR limit.
The idea consists in introducing Lorentz breaking terms in the action such that the possibility appears that the resulting theory becomes renormalisable in the UV limit. 
Indeed, it was established that
if space and time re-scale differently as $x^{i}\rightarrow lx^{i},\quad t\rightarrow l^{z}t$, where $z$ is the scaling exponent, then in the case $z=3$ one can add non-relativistic terms to the Einstein-Hilbert action in such a way that  the resulting theory is renormalisable by power counting. 
The problem appears that the most general action for such a theory contains about 100 terms, which makes the problem quite difficult to handle and drastically reduces the predictability of the theory due to the large number of arbitrary coupling constants entering the action \cite{wang17}. It is, therefore, convenient to consider particular cases with a reduced number of coupling constants. To this end, additional symmetries are imposed at the level of the action, which often lead to new difficulties that affect either the energy limits or even the renormalisability of the theory. 
Several actions have been proposed so far in order to solve different conceptual problems that appear at the UV and IR energy levels.  
In the original HL model, two major simplifying conditions were assumed, namely, projectability and  detailed balance which, however, lead to difficulties \cite{cnps09,bps09}. The first assumption led to inconsistencies in the IR limit, which can be  
treated by imposing strong conditions on the free parameters of the general action
\cite{Sotiriou:2009gy}. In addition, instabilities at the UV and the IR energetic limits
were found \cite{Mukohyama:2010xz,Izumi:2011eh,Gumrukcuoglu:2011ef}  that make it difficult to handle these limits from a physical point of view, even at the level of particular gravitational configurations \cite{Mukohyama:2010xz,Izumi:2011eh,Gumrukcuoglu:2011ef}. Although, several results concerning instabilities are controversial \cite{wang17}, the simplest solution to this problem is to consider only non-projectable configurations  \cite{Fukushima:2018xgv}.

Despite these difficulties and  ambiguities, HLG has been applied in many frameworks, especially in black hole physics and cosmology \cite{ Ramos:2018oku,Ma:2016lwr,Gao:2020ddm,Nilsson:2019bxv}.
In this work, we will consider one of the simplest versions of HLG, in which the conditions of projectability and detailed balance are imposed \cite{hor09a}. 
The original version of this minimal model had the problem that the Schwarzschild-AdS black hole metric was not a solution of the field equations in the corresponding IR limit.  This problem was solved by including ad additional term in the action that leads to IR field equations with the Schwarzschild-AdS spacetime as a particular solution \cite{park09}, implying at the same time to a redefinition of the IR limit. 
As a result, only six coupling constants enter the action  that can be written as
\color{black}
\begin{eqnarray}\label{eq:action_Horava}
S_g& =& \int d^4 x N \sqrt{g}  \left[\frac{2}{\kappa^2}\left(K_{ij}K^{ij}-\lambda K^2\right)-\frac{\kappa^2}{2\nu^4}C_{ij}C^{ij}
+
\frac{\kappa^2\mu}{2\nu^2}\epsilon^{ijk} R^{ }_{i\ell} \nabla_{j}R^{ \ell}{}_k -\frac{\kappa^2\mu^2}{8} R^{ }_{ij} R^{ ij} +
\right. \nonumber \\
&&  \left.
 \frac{\kappa^2 \mu^2}{8(3\lambda-1)}
\left(\frac{4\lambda-1}{4}R^2-\Lambda_W R+3\Lambda_W^2\right)+
 \frac{\kappa^2 \mu^2\omega}{8(3\lambda-1)} R\right],
\end{eqnarray}
where
\begin{eqnarray}
K_{ij}&=&\frac{1}{2N} (\dot g_{ij}-\nabla_i N_j -\nabla_j N_i)\,, \\
C^{ij}&=& \epsilon^{ikl} \nabla_k \left(R^j_l -\frac{1}{4} R \delta^j_l\right)\,,
\end{eqnarray}
\noindent are the extrinsic curvature and the Cotton tensor, respectively. The dot represents the derivative with respect to the time coordinate, and $R$ is the scalar curvature. Further, $N$ and $N_i$ are the lapse and shift in the 3+1 decomposition, i.e.
$ds^2 = -N^2 c^2 dt^2 + g_{ij}(dx^i + N^idt)(dx^j+N^jdt)$.

This model presents a significant degeneracy since it depends upon six parameters:
$\kappa$, $\lambda$, $\mu$, $\nu$, $\Lambda_W$ and $\omega$, which are not \textit{completely} free. Some of them
determine the speed of light $c$, the gravitational constant $G$ and the cosmological constant $\Lambda$ according to
\begin{eqnarray}
\mu^2 &=& \frac{16 c^2 (3\lambda -1)^2 |\Lambda|}{3\kappa^4 \Lambda^{2}_{W}},  \label{eq:mu} 
\quad G=\frac{\kappa^2c^2}{16\pi(3\lambda-1)}\ , \label{eq:G} 
\end{eqnarray}
where $\Omega_{\Lambda_W} = \frac{2}{3}\Omega_\Lambda\,$, and if $\lambda<1/3$ implies the presence of repulsive gravitational effects. The contrary case ($\lambda>1/3$) leads to attractive gravity. 
In the IR limit, the HL action reduces to 
\begin{equation}
S_{IR} = \frac{c^2}{8\pi G (3\lambda -1)} \int d^4 x \sqrt{g} N \Bigg[ K_{ij} K^{ij} - \lambda K^2 
+ c^2 R \left(1 + \frac{3\omega c^2 (3\lambda -1)}{ 4 |\Lambda|} \right)\Bigg]\ ,
\end{equation}
which is expected to be equivalent to the GR action. Then, it follows that the Einstein-Hilbert action with cosmological constant is obtained in the limit $\lambda\to 1$ and $\omega\to 0$. The limit $\lambda \to 1$ is also consistent with the expected values of the universal constants $G$, $c$, and $\Lambda$. We will see below, in Sec. 7, that the constant $\omega$ encodes completely the quantum contribution of the HL gravity model at the cosmological level. Consequently, the IR limit $\omega\to 0$ is also consistent with this interpretation. 

In this work, we will study the large-scale cosmological model described by the Friedmann-Lema\^itre-Robertson-Walker (FLRW) metric
\begin{equation}
ds^2 = - c^2 dt^2 + a^2(t)\left[ \frac{ dr^2}{1-k\frac{r^2}{r_0^2}} + r^2 (d\theta^2 + \sin^2 \theta d\varphi^2)\right] \ .
\end{equation}
The use of this metric could arouse certain concerns. Indeed, 
when considering perturbations 
around a background metric, as mentioned above, it has been argued that HLG does not have an IR limit corresponding to GR because of the presence of strong coupled gravity fluctuations. This would be a major problem for the consistency of the theory. Nevertheless, a detailed discussion of this problem is beyond the scope of the present
work. In practice, however, this problem might not be quite relevant to our case since we will consider only cosmological solutions with a non-vanishing but tiny cosmological constant, as favored by current observations. In fact, when considering perturbations around such cosmological solutions a natural cut-off appears due to the tiny value of the cosmological constant. Of course, a more detailed analysis will be necessary to confirm this intuitive idea.

Under the ansatz $\Lambda_W > 0$, the only scenario to be in agreement with local scale limits, and considering a 
perfect-fluid source with energy density and pressure $\rho$ and $p=w\rho$, respectively, we can compute the new Friedmann equations for this theory and obtain 
\begin{eqnarray}\label{eq:friedmann}
\left(\frac{\dot{a}}{a}\right)^2&=&b_1\Bigg\{\rho \pm b_2 \left[-\Lambda^2_W + \frac{2k(\Lambda_W-\omega)}{r^2_0a^2}-\frac{k^2}{r^4_0a^4}\right]\Bigg\},\quad\quad \\
\frac{\ddot a}{a}&=&b_1\left[-\frac{1}{2}(\rho+3p) \pm b_2 \left(-\Lambda^2_W+\frac{k^2}{r^4_0a^4}\right)\right]\,,
\end{eqnarray}
with $b_1=\kappa^2/[6(3\lambda-1)]$ and $b_2=3\kappa^2\mu^2/[8(3\lambda-1)]$.  We can combine the latter expression and write a differential equation in terms of the redshift $a=(1+z)^{-1}$ as
the corresponding differential equation, obtained by combining the above two expressions, can be recast as
\begin{eqnarray}
&&(1+z)\frac{dH^2}{dz}-3(1+w)H^2+\Lambda(1+w)c^2 
-\frac{\Lambda kc^2}{3\Lambda_{W}^{2}r_{0}^{2}}
\Big[(1+3w)(\Lambda_W-\omega)(1+z)^2  
\nonumber \\ &&
+\frac{k}{2r_{0}^{2}}(1-3w)(1+z)^4\Big]=0\,. \quad
\label{feq}
\end{eqnarray}
We defined the Hubble rate $H\equiv \dot{a}/a$ and $|\Lambda| = 3\mu^2\kappa^4\Lambda^2_W /16c^2(3\lambda-1)^2$.
Following, we need to characterise the type of fluid in the HLG to study its dynamics.

Equation (\ref{feq}) shows that the term proportional to $a^{-4}$ does not appear in the corresponding model in Einstein gravity and so it represents the contribution of HLG. This term vanishes for $k=0$ or $w=1/3$. This indicates that in the case of a flat universe with arbitrary EoS or a radiation dominated universe with arbitrary topology, there is no difference between Einstein gravity and HLG. Since in this work we are interested in analyzing HLG in the lowest redshift regime, i.e., the recently observed universe, we can neglect the contribution of radiation and consider a fluid with $w=0$ as the next interesting case. Indeed, this pressureless fluid can be used to describe either baryonic matter or cold dark matter, which together with the cosmological constant constitute the main components of the Universe. 
Accordingly, we will consider now the case of dust, which is also the simplest case of a matter term with EoS: $p=0 \ (w=0)$.

In \cite{Luongo:2018oil}, the Ho\v rava corrections to the Friedmann equations were found, showing that they become relevant for different regimes, e.g.,  at the low redshift regime, the term $\omega$ does not contribute so that it is difficult to bound it in a FLRW scenario. If we assume a constant barotropic factor $w$ for a given cosmological fluid, the Hubble rate $H(z)$ can be written as
\begin{eqnarray}\label{eos_horava}
E^2(z)&=&\sum_i\Omega_i(1+z)^{3(1+w_i)}+\Omega_k\left(1-\frac{\Omega_\omega}{\Omega_\Lambda}\right)(1+z)^2 
+\Omega_{r}^{*}(1+z)^4+\Omega_\Lambda\,,
\end{eqnarray}
where $E^2(z)=H^2/H_0^2$ and $|\Omega_\omega|\equiv\omega c^2/(2H_0^2)$. 
\\


\section{Ho\v rava-Lifshitz inverse cosmography} 

To compute the \textit{generic} cosmograhic parameters for the HLG theory, we use its EoS (\ref{eos_horava}) in the inverse cosmography equations (\ref{eq:q})-(\ref{eq:j})-(\ref{eq:s}). The simplest scenario to be considered is the HLG dust case:
\begin{eqnarray}
q&=&-1+ \frac{(z+1)^2 \left(\Omega _k \left(2-\frac{2 \Omega _w}{\Omega _{\Lambda }}\right)+(z+1) \left(3 \Omega
   _m+4 (z+1) \Omega^{*}_{r}\right)\right)}{2 \left((z+1)^2 \Omega _k \left(1-\frac{\Omega
   _w}{\Omega _{\Lambda }}\right)+\Omega _{\Lambda }+(z+1)^3 \Omega _m+(z+1)^4 \Omega^{*}_{r}\right)} \biggr |_{z=0} 
   \nonumber\\
   &=&  -1+ \frac{\Omega _k \left(2-\frac{2 \Omega _w}{\Omega _{\Lambda }}\right)+3 \Omega _m+4 \Omega^{*}_{r}}{2 \left(\Omega _k \left(1-\frac{\Omega _w}{\Omega _{\Lambda }}\right)+\Omega _{\Lambda
   }+\Omega _m+\Omega^{*}_{r}\right)},  \quad \quad 
  \\
 j&=& \frac{\Omega _{\Lambda } \left(\Omega _{\Lambda }+(z+1)^3 \Omega _m+3 (z+1)^4 \Omega^{*}_{r}\right)}{\Omega _{\Lambda } \left(\Omega _{\Lambda }+(z+1)^3 \Omega _m+(z+1)^4 \Omega^{*}_{r}\right)-(z+1)^2 \Omega _k \left(\Omega _w-\Omega _{\Lambda }\right)} \biggr |_{z=0} 
  \nonumber\\
& =& \frac{\Omega _{\Lambda } \left(\Omega _{\Lambda }+\Omega _m+3\Omega^{*}_{r}\right)}{\Omega
   _{\Lambda } \left(\Omega _{\Lambda }+\Omega _m+\Omega^{*}_{r}\right)-\Omega _k \left(\Omega
   _w-\Omega _{\Lambda }\right)},
 \\
   s&=& \frac{\Omega _{\Lambda } \left(\Omega _{\Lambda }+z (z+1)^2 \Omega _m+(3 z-1) (z+1)^3 \Omega^{*}_{r}\right)}{\Omega _{\Lambda } \left(\Omega _{\Lambda }+(z+1)^3 \Omega _m+(z+1)^4 \Omega^{*}_{r}\right)-(z+1)^2 \Omega _k \left(\Omega _w-\Omega _{\Lambda }\right)},  \biggr |_{z=0} 
    \nonumber\\
&= & \frac{\Omega _{\Lambda } \left(\Omega _{\Lambda }-\Omega^{*}_{r}\right)}{\Omega _{\Lambda }
   \left(\Omega _{\Lambda }+\Omega _m+\Omega^{*}_{r}\right)-\Omega _k \left(\Omega _w-\Omega
   _{\Lambda }\right)},
   \\
   l&=& \frac{\Omega _{\Lambda } \left(\Omega _{\Lambda }+z (z+1)^2 \Omega _m+(z (3 z+2)+1) (z+1)^2\Omega^{*}_{r}\right)}{\Omega _{\Lambda } \left(\Omega _{\Lambda }+(z+1)^3 \Omega _m+(z+1)^4 \Omega^{*}_{r}\right)-(z+1)^2 \Omega _k \left(\Omega _w-\Omega _{\Lambda }\right)} \biggr |_{z=0} 
    \nonumber\\
   &=& \frac{\Omega _{\Lambda } \left(\Omega _{\Lambda }+\Omega^{*}_{r}\right)}{\Omega _{\Lambda }
   \left(\Omega _{\Lambda }+\Omega _m+\Omega^{*}_{r}\right)-\Omega _k \left(\Omega _w-\Omega
   _{\Lambda }\right)}.
\end{eqnarray}
\\
We can derive expression of HLG parameter $\Omega_W$ in terms of the cosmographic parameters as at $z=0$ denoted by the subindex \textit{0}:
\begin{eqnarray}
\Omega_W (q_0) &=& \frac{\Omega _{\Lambda } \left(2 q_0 \Omega _k+2 \Omega _{\Lambda }+2 q_0 \Omega _m-\Omega
   _m+2 q_0 \Omega _{\Lambda }\right)}{2 q_0 \Omega _k}, ~\quad \\
   \Omega_W (j_0) &=& \frac{\Omega _{\Lambda } \left(j_0 \Omega _k+j_0 \Omega _{\Lambda }+j_0 \Omega
   _m-\Omega _{\Lambda }-\Omega _m\right)}{j_0 \Omega _k}, \\
     \Omega_W (s_0) &=& \frac{\Omega _{\Lambda } \left(s_0 \Omega _k-\Omega _{\Lambda }+s_0 \Omega _m+s_0
   \Omega _{\Lambda }\right)}{s_0 \Omega _k}.
\end{eqnarray}
In Figure 1, we illustrate the cosmographic parameters for Ho\v rava in comparison to standard dust case in an observational redshift range of interest.

\begin{figure*}
\centering
    \includegraphics[width=0.45\textwidth,origin=c,angle=0]{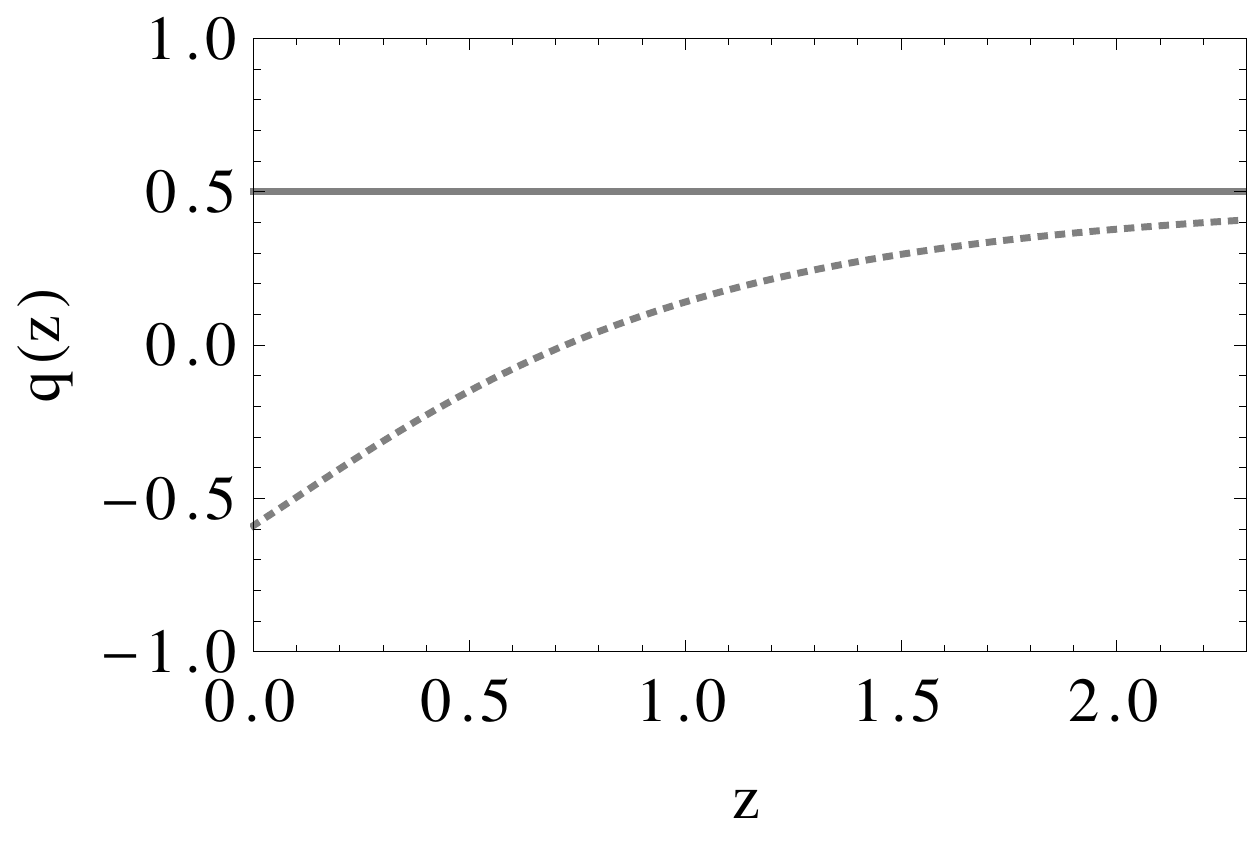}
          \includegraphics[width=0.47\textwidth,origin=c,angle=0]{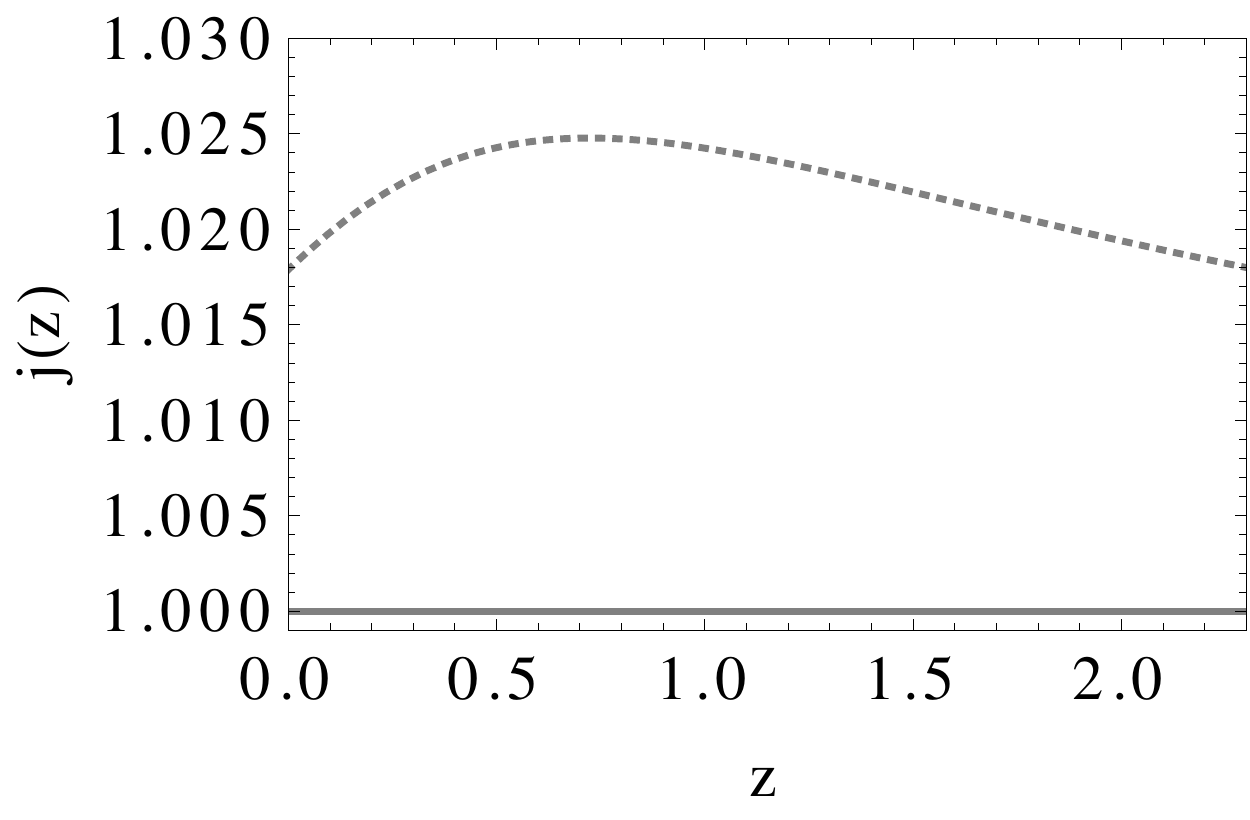}
              \includegraphics[width=0.45\textwidth,origin=c,angle=0]{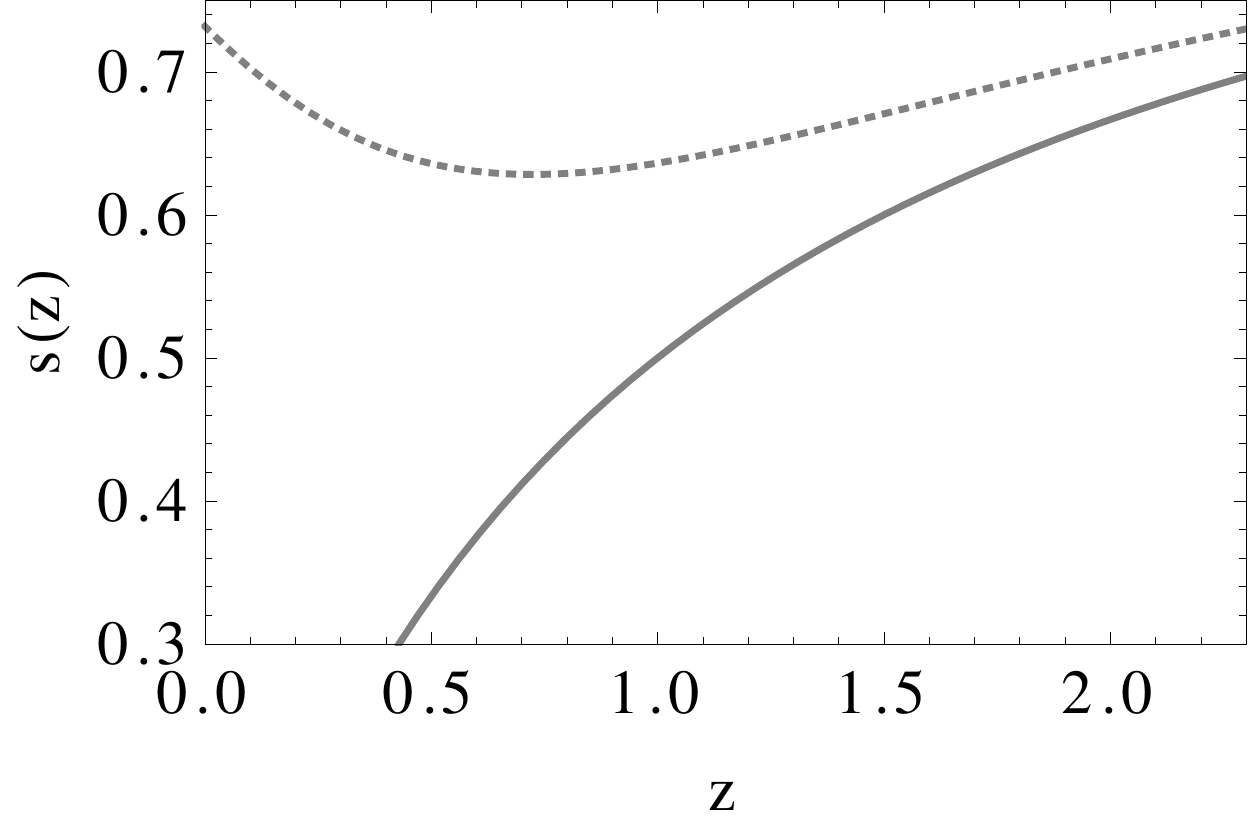}
    \caption{Cosmographic parameters for Ho\v rava (dashed line) in comparison to standard dust case (solid line).}
    \label{fig:cosmography_Horava}
\end{figure*}

\begin{figure}
\centering
    \includegraphics[width=0.7\textwidth,origin=c,angle=0]{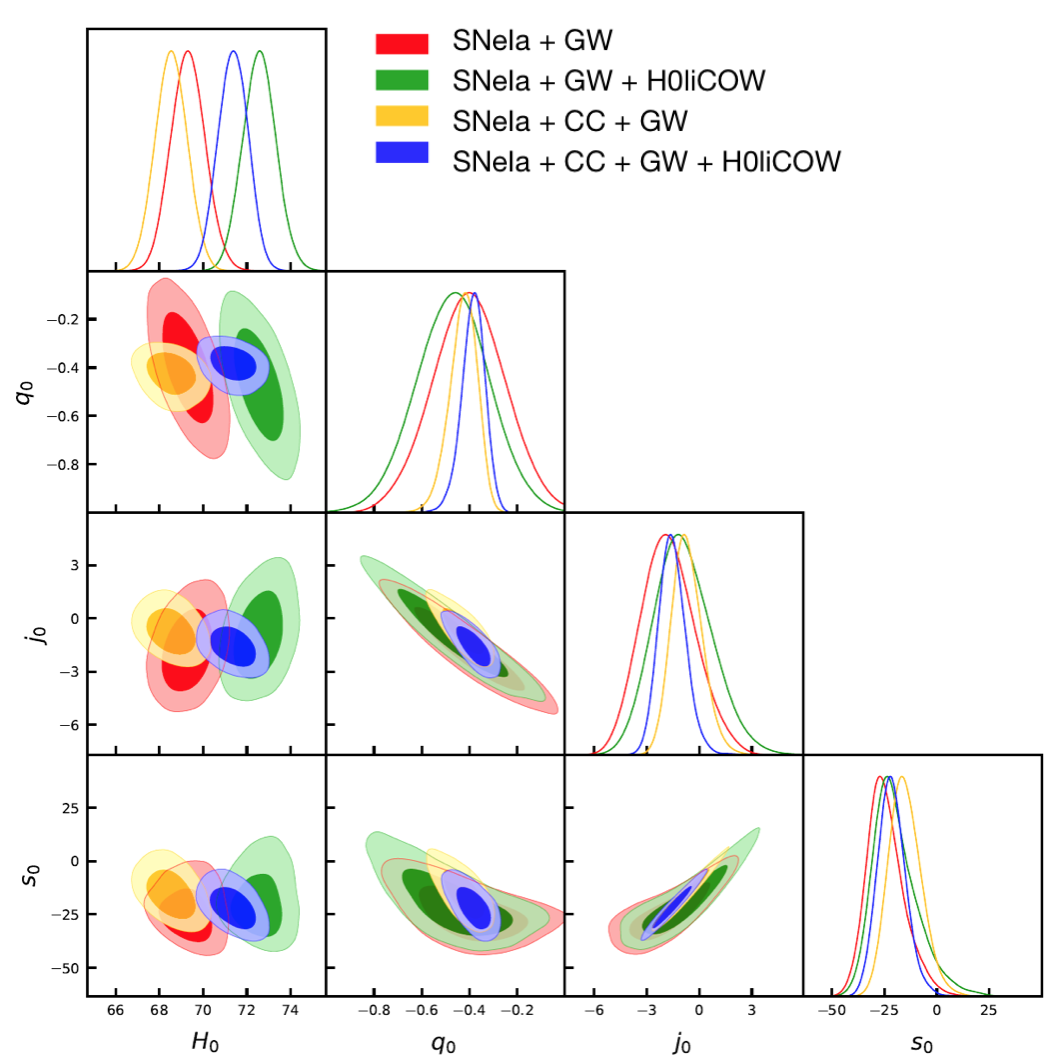}
    \caption{C.L Cosmographic parameters space for Ho\v rava dust model at $z=0$ using observational samplers: Pantheon, GW, H0liCOW and CC.}
    \label{fig:cosmography_Horava}
\end{figure}

\section{Observational and forecast surveys}

\begin{enumerate}
\item Pantheon SNeIa sampler: Consist 
of 1048 SN Ia in a redshift region $z \in [0.01, 2.3]$ \cite{Scolnic:2017caz}, whose distance moduli are standardized through the SALT-2 light-curve fitter.
Under the only assumption of a flat universe, the full Pantheon catalog can be compressed into six model-independent $E^{-1}(z)$ measurements. Therefore, we use these measurements correlated among them according to the covariance matrix $C_{ij}$. In this case, the likelihood probability function is described by
\begin{equation}
\mathcal{L}_{SNeIa} \propto \exp \left( -\frac{1}{2}\bm{V}^{T} C_{ij}^{-1}\bm{V} \right) ,
\end{equation}
where $\bm{V}=E_{obs}^{-1}-E_{th}^{-1}$ is the differences between the observed and the theoretical expectation values.

\item Cosmic Chronometers: The late cosmic expansion can be studied in a model-independent way by measuring the age difference of Cosmic Chronometers (CC). 
\cite{Jimenez:2019onw}. 
From the spectroscopic measurements of the redshifts between pairs of  old and passively evolving galaxies and their differential age, one can obtain the Hubble parameter
$H(z)=-(1+z)^{-1}dz/dt $.
In this work we consider 31 uncorrelated measurements of $H(z)$ in a redshift range $z: [0,2]$. Confronting these values with the corresponding Hubble expansion rates predicted by the theoretical scenarios, one can write the likelihood function as
\begin{equation}
\mathcal{L}_{CC}\propto \exp{\left[-\frac{1}{2}\sum_{i=1}^{31}\left(\frac{H^{obs}_i-H^{th}_i}{\sigma_{H,i}}\right)^2\right]}.
\end{equation}

\item H0LiCOW data: For this sampler, we employ the time-delay distance given by
\begin{equation}
D_{\Delta t}=(1+z_l)\frac{D_l D_s}{D_{ls}}\	 , 
\end{equation}
where $z_l$ is the redshift of the lens and
\begin{eqnarray}
D_l&=& \frac{H_0^{-1}}{1+z_l}\int_0^{z_l}\frac{dz}{E(z)}\ , \\
D_s&= & \frac{H_0^{-1}}{1+z_s}\int_0^{z_s}\frac{dz}{E(z)}\ , \\
D_{ls}&=&\frac{H_0^{-1}}{1+z_s}\int_{z_l}^{z_s}\frac{dz}{E(z)}\ ,
\end{eqnarray}
and $z_s$ are the angular diameter distances to the lens, to the source, between the lens and the source and the source redshift, respectively.
In this work, we use the six systems of strongly lensed quasars analysed by the H0LiCOW collaboration \cite{Rusu:2019xrq}, in where the
likelihood function for the $D_{\Delta t}$ is described by 
\begin{equation}
\mathcal{L}_{H0LiCOW}\propto \exp\left[-\frac{1}{2}\sum_{i=1}^{6}\left(\frac{D_{\Delta t, i}^{obs}-D_{\Delta t,i}^{th}}{\sigma_{D_{\Delta t},i}}\right)^2\right] .
\end{equation}

\item Standard Sirens ET simulations: We use here simulations from a third-generation
ground-based GW detector, Einstein Telescope (ET), since it will allow to have a better statistical sensitivity
detection in comparison to the current LIGO detectors. The advantage of this observation is that it can be used
to relax (or even break) degeneracies generated by other surveys due to their definition in terms of the absolute luminosity distance \cite{Kyutoku:2016zxn}, and 
its redshift range will reach $z\approx 2-3$.
To perform the simulation, we first consider the best fits parameters for the HLG dust model obtained from H0LICOW + Pantheon + CC, then this will be the fiducial model to generate the GW standard sirens simulations from the ET\footnote{The full method is described in \cite{Zhang:2018byx}.}. The simulation consist in 1000 standard sirens from the ET, where each point consist of a vector given by redshift, the luminosity distance and its error as: $(z_i, d_{L}, \sigma_i )$. The set of simulated GW data will consist of $N$ data points and the standard likelihood can be described by
\begin{equation}
\mathcal{L}_{GW}\propto \exp \left(-\frac{1}{2}\sum^{N}_{i=1} \left[\frac{d^{i}_{L}- d_{L}(z_{i};\Theta)}{\sigma^{i}_{d_L}}\right]^2\right) ,
\end{equation}
where $z_i$, $d^{i}_{L}$ and $\sigma^{i}_{d_L}$ are the $i$th redshift, luminosity distance and its error, respectively. $\Theta$ denotes the set of HLG cosmographic parameters obtained by the above samplers.
\end{enumerate}

{\renewcommand{\tabcolsep}{0.03mm}
{\renewcommand{\arraystretch}{0.6}
\begin{table}                   
\caption{Cosmographic parameters results for HLG theory dust case with SNeIa Pantheon, CC, H0LICOW, GW and the combinations of the samplers.}                                                                                               
\begin{tabular}{c|ccccc}                                                                                                            
\hline                        
&& &  Best fits         \\                                                                                   
\small{Parameters} & Pantheon & CC & GW & \small{H0LiCOW} & \small{SN + CC + GW + H0LiCOW} \\ \hline \\
$H_0$ & $   71.84_{-    0.708 }^{+    0.706}$ & $69.010_{-    1.332}^{+    1.329}$  & $   72.574_{-    0.757}^{+    0.747}$ & $   71.367_{-    0.689}^{+    0.6840}$ &$   71.904_{-    1.347 }^{+    1.406}$\\ \\
$q_0$ & $   -0.424_{-    0.060}^{+    0.077}$ &  $   -0.533_{-    0.119}^{+    0.142}$ & $   -0.450_{-    0.152}^{+    0.152}$& $   -0.385_{-    0.045}^{+    0.053}$ &$   -0.423_{-    0.141}^{+    0.133}$\\ \\
$j_0$ & $   1.199_{-    1.003}^{+    0.907}$ &  $  0.114_{-    1.459}^{+    1.322}$ &$   0.912_{-    1.832}^{+    1.518}$  & $   1.539_{-    0.834}^{+    0.704}$ &$   1.210_{-    1.863}^{+    1.891}$\\ \\
$s_0$ & $  -18.484_{-    8.825}^{+    7.945}$  & $  -11.171_{-   13.405}^{+   11.101}$ & $  -19.428_{-   12.790}^{+    7.509}$ & $  -21.387_{-    7.255}^{+    6.164 }$ &  $  -18.510_{-    16.514}^{+    16.746}$\\ \\
\hline                                                                                                                
\end{tabular}                                                                                                                   
\label{tab:results2}                                                                                                   
\end{table}}}                                                                                                                  
                                                                                                                 
{\renewcommand{\tabcolsep}{0.03mm}
{\renewcommand{\arraystretch}{0.6}
\begin{table}      
\caption{Best fits values for HLG theory parameters SNeIa Pantheon, CC, H0LICOW, GW and the join sampler.}                                                                                                             
\begin{tabular}{c|ccccc}                                                                                                            
\hline                        
&& & &Best fits         \\                                                                                   
Parameters & Pantheon & CC & GW & H0LiCOW &SN + CC + GW + H0LiCOW \\ \hline \\
$\Omega_m$ & $0.292^{+0.032}_{-0.132}$   & $0.311^{+0.022}_{-0.013}$   & $0.325^{+0.072}_{-0.252}$	& $0.308^{+0.023}_{-0.013}$ & $0.331^{+0.101}_{-0.101}$\\ \\
$\Omega_{\Lambda}$ 	&$0.718^{+0.706}_{-0.708}$	&$0.690_{-    1.332}^{+    1.329}$ 	&$   0.726_{-    0.757}^{+    0.747}$ &$   0.714_{-    0.689}^{+    0.6840}$ & $   0.719_{-    1.347 }^{+    1.406}$  \\ \\
$\Omega_k$ & $-0.012^{+0.12}_{-0.18}$ & $-0.012^{+0.02}_{-0.02}$ & $-0.020^{+0.013}_{-0.019}$& $-0.01^{+0.021}_{-0.018}$ & $-0.021^{+0.023}_{-0.029}$\\ \\
$\Omega_{\omega}$ & $0.090^{+4.01}_{-4.03}$ & $0.081^{+3.11}_{-3.13}$ & $0.120^{+3.18}_{-3.18}$& $0.122^{+5.01}_{-4.12}$ & $0.090^{+3.241}_{-3.241}$\\ \\
\hline                                                                                                                
\end{tabular}                                                                                                                   
\label{tab:results1}                                                                                                   
\end{table}}}                                                                                                                      
                                                                                                                  
{\renewcommand{\tabcolsep}{0.03mm}
{\renewcommand{\arraystretch}{0.6}
\begin{table}                
\caption{Derived quantities values for HLG theory using the results given in Table \ref{tab:results2}.}                                                                                                       
\begin{tabular}{c|ccccc}                                                                                                            
\hline                        
Parameters & Pantheon & CC & GW & H0LiCOW & SN+CC+GW+H0LiCOW \\ \hline \\
$\Omega_{\Lambda_\omega}$ 	& $0.195^{+0.021}_{-0.008}$ 	& $0.207^{+0.015}_{-0.009}$ 	&$0.217^{+0.048}_{-0.168}$ &$0.205^{+0.015}_{-0.009}$ &$0.220^{+0.067}_{-0.067}$\\ \\
$\omega \times 10^{-14} L^{-2}$ & $-3.349^{+331.128}_{-1363.33}$ &$-3.291^{+211.198}_{-125.479}$ &$-3.804^{+759.417}_{-2655.54}$& $-3.386^{+235.127}_{-133.273}$ & $-3.803^{+1046.24}_{-1046.17}$\\ \\
$\Omega_{W} (q_0)$ 	& $21.005^{+10.021}_{-10.008}$ 	& $0.766^{+1.015}_{-1.009}$ 	&$8.030^{+1.048}_{-0.148}$ &$31.598^{+12.115}_{-12.009}$ &$9.600^{+2.167}_{-2.167}$\\ \\
$\Omega_{W} (j_0)$ 	& $9.311^{+2.021}_{-1.008}$ 	& $454.279^{+322.115}_{-309.109}$ 	&$4.40^{+3.148}_{-3.168}$ &$24.842^{+11.115}_{-8.933}$ &$5.520^{+2.667}_{-2.667}$\\ \\
$\Omega_{W} (s_0)$ 	& $62.038^{+32.021}_{-28.008}$ 	& $61.224^{+23.015}_{-19.449}$ 	&$38.789^{+12.148}_{-12.148}$ &$74.641^{+13.015}_{-7.009}$ &$36.561^{+10.133}_{-5.121}$\\ \\
\hline                                                                                                                
\end{tabular}                                                                                                                   
\label{tab:results3}                                                                                                   
\end{table}}}                                                                                                                 

\section{HLG quantum signatures} 

To perform the cosmographic statistics, we employ combinations from the observations described and
denote the total $\chi^{2}_{tot}$ function as
\begin{equation}
\chi^2_{tot} = \chi^{2}_{Pantheon} +\chi^{2}_{CC} +\chi^{2}_{H0LICOW} +\chi^{2}_{GW}.
\end{equation}
Our best fits values for the HLG theory parameters are reported in Table \ref{tab:results1}. For the cosmographic parameters derived from the HLG theory in the dust case scenario are reported in Table \ref{tab:results2}. Finally, in Table \ref{tab:results3} we present the values for the derived quantities for the HLG theory using the results reported in the latter Tables. In Figure \ref{fig:cosmography_Horava} we present the cosmographic parameters space for Ho\v rava dust model at $z=0$ using observational samplers described above.

From (\ref{eq:friedmann}), we notice that $a^{-4}$ denotes the contribution from the higher-derivatives terms that are present in the Ho\v rava action and are expected to play a relevant role in the UV limit. This term vanishes in the case of a spatially flat universe $(k=0)$, implying that UV contributions can be detected only if $k\neq 0$. We interpret this result as an indication that the quantum effects in HLG cosmology necessarily lead to a spatially curved universe. This is probably the reason why a comparison of HLG cosmology with the flat $\Lambda$CDM model leads to inconsistencies \cite{Luongo:2018oil}. To evaluate the contribution of the quantum effects in the HLG cosmology as given by (\ref{eq:friedmann}), we introduce the notation 
\begin{equation}
\rho_k \equiv \frac{2 k (\Lambda_W-\omega)}{r^2_0a^2},\, \quad \rho_Q\equiv \frac{k^2}{r^4_0a^4}.
\end{equation}
Then, the quantum contribution $\rho_Q$ is related to the curvature contribution $\rho_k$ by 
\begin{equation}
\frac{\rho_k^2}{\rho_Q}= 4 (\Lambda_W - \omega)^2\ .
\end{equation}
Afterward, we can compute the constant $\Lambda_W-\omega$ by using our samplers and this will allow us to estimate the contribution of the quantum effects at the cosmological level. Recall that $\Lambda_W$ represents in this theory the cosmological constant. Therefore, the quantum contribution is entirely contained in the constant $\omega$.

In particular, for the best fits obtained using the join samplers Pantheon+CC+GW+H0LiCOW, we can set current bounds at cosmic scales in order to found a quantum signature.
To this end, we evaluate the quantum contribution 
\begin{equation}\label{eq:qc}
 \frac{\rho^{2}_{k}}{\rho_Q}= 5.785 \times 10^{-27} m^{-2},
\end{equation}
where we used the value for the cosmological constant $\Lambda = 1.1056\times 10^{-52} m^{-2}$ given by Planck 2018 and our best fit for $\omega$ for the join sampler reported in Table \ref{tab:results3}. This result is crucial since we are obtaining a direct value of the HLG parameter $\omega \approx -3.8\times 10^{-14}$, from late time observations that have effect in the density quantum contribution (\ref{eq:qc}). These effects will have negative sign contributions if we deal with the modified Friedmann constriction $\Omega_{m} +\Omega_{\Lambda} +\Omega_{Q} +\Omega^{*}_{r}=1$, with
$\Omega_{Q}\equiv \Omega_k \left(1- \frac{\Omega_\omega}{\Omega_{\Lambda}}\right), \quad \Omega^{*}_{r}\equiv \frac{\Omega^{2}_{k}}{4\Omega_{\Lambda}}+\Omega_{r}, $ since the HLG terms are related to the curvature density parameter that need to be compensated by the total matter/energy contributions with the values of the parameters $b_1$ and $b_2$.

It is interesting to note that in the IR limit $\omega \to 0$, we obtain values for the cosmographic parameters which agree with the standard $GR$ values for a non-flat cosmology. This works as a smocking gun analysis in order to check the viability of our proposal.


\section{Discussion} 

In this paper we present a proposal of inverse cosmography in order to set quantum signatures on Ho\v rava gravity via its cosmographic parameters, without experimenting truncation problems over the Taylor series. With this approach we obtain a HLG generic cosmography scenario, where it is possible to study parameters that came from quantum signatures when a non-flat universe is considered. Notice that this kind of non-flat scenario also has been considered in \cite{Handley:2019tkm,DiValentino:2019qzk}, to explain the presence of a lensing amplitude in CMB linear power spectra in comparison to the standard $\Lambda$CDM, which report a value of $\Omega_k = 0.0007\pm 0.0019$ at $68\%$ CL. Even though, this latter result relaxes the $H_0$ tension problem by estimated a value of $\Omega_k =-0.091\pm 0.037$ at $68\%$ C.L., i.e a closed universe. The problem with this result lies in adding curvature as an additional free parameter to the standard six from $\Lambda$CDM (so-called $\Lambda$CDM+$\Omega_k$) and a deviation of around 3-$\sigma$ arise with respect to Planck 2018 prediction and the BAO observations in different galactic catalogs. With our HLG proposal, we obtain \textit{naturally} a non-flat scenario with $\Omega_k =-0.021^{+0.023}_{-0.029}$ with $H_0 = 71.904^{+1.406}_{-1.347}$, without showing a $3.4$-$\sigma$ inconsistency as it was reported in the latter reference. Also, we employ forecast GW data from Einstein Telescope to enrich the observational redshift range of the other samplers. Moreover, according to \cite{Efstathiou:2020wem}, including BAO sample leads to a strong preference for a flat universe, wherever the CMB likelihood is used. Therefore, to relax this bias, we consider only distance ladder measurements in this work.

The main result of the present work is that by applying the novel method of inverse cosmography in the HLG cosmological model, we can explicitly analyse the contribution of the terms that are relevant in the UV regime, i.e., that are related to the quantum nature of the model.
Notice, however, that we are not testing the HLG model in the UV regime. Indeed, the additional terms  of the HL action lead to the modified Friedman equations, which are then evaluated and confronted with observations in the IR regime. That is, although the additional HL terms are expected to be essential in the UV regime and small in the IR limit, we investigate the possibility that they lead to measurable effects in the modified Friedman equations, according to current observations. 
 We obtain as a first result that our universe must be  spatially curved due to the presence of the additional HL terms in the IR limit. This  implies that a different inflationary scenario would be necessary to be in agreement with a spatially curved universe.

  Since in Einstein gravity, the $\Lambda$CDM model predicts a flat universe, 
{ we can conclude that classically our universe is flat, but from our results we notice that in this HLG scenario can be curved 
due to the presence of additional terms in the HL action.}
Moreover, by using recent results of cosmological and Gravitational Wave observations, we were able to obtain a concrete value for the Ho\v rava parameter  $\omega \approx -3.8\times 10^{-14}$. This is a concrete result of our analysis that  could be compared with results obtained in other scenarios in order to establish the observational validity of HLG at the quantum level.  

To perform the cosmographic analysis presented in this work, we have chosen  an explicit  cosmological scenario in a particularly simple HLG model, which has been the subject of intensive criticism regarding stability. We argue that the presence of a tiny cosmological constant could be used as a natural cut-off that prevents instabilities. However, the confirmation of this idea would imply an extensive analysis, which is beyond the scope of the present work. Indeed, the IR limit at the level of the HL action is consistently obtained if the running parameters as chosen as $\lambda\to 1$ and $\omega\to 0$. Stability problems have been found in the limit $\lambda\to 1$. If the additional parameter $\omega$ could be interpreted as corresponding to a supplementary degree of freedom, it would necessary to perform a detailed analysis to see if this could meliorate the stability problem. Another possibility to avoid from the very beginning the problem of instabilities would be to consider the general non-projectable action and probably also without the detailed balance condition. Again, this task is beyond the scope of the present work, but we expect to explore in future investigations.


\section{Acknowledgments}\label{sec:acknowledgements} 

CE-R acknowledges the Royal Astronomical Society as FRAS 10147 and the support 
by PAPIIT Project IA100220 and would like to acknowledge networking support by the COST Action CA18108.
This work was partially supported  by UNAM-DGAPA-PAPIIT, Grant No. 114520, 
Conacyt-Mexico, Grant No. A1-S-31269, 
and by the Ministry of Education and Science (MES) of the Republic of Kazakhstan (RK), Grant No. 
BR05236730 and AP05133630.



\section*{References}


\begin{thebibliography}{0} 

\bibitem{carlip01}
S. Carlip, Rept. Prog. Phys. {\bf 64}, 885 (2001).

\bibitem{kiefer05}
 C. Kiefer, Annal. Phys.  {\bf 15}, 129 (2005).
 
 
 \bibitem{hor09a}
 P. Ho\v rava,  Phys. Rev. D, {\bf 79}, 084008, (2009).
 

 \bibitem{hor09b}
 P. Ho\v rava,  Phys. Rev. Lett., {\bf 102}, 161301, (2009).
 
 \bibitem{Luongo:2018oil} 
 O.~Luongo, M.~Muccino and H.~Quevedo,
 Phys.\ Dark Univ.\  {\bf 25}, 100313 (2019)
 doi:10.1016/j.dark.2019.100313
 [arXiv:1811.05227 [gr-qc]].
 
 \bibitem{Munoz:2020gok}
 C.~Z.~Munõz and C.~Escamilla-Rivera,
 Accepted by JCAP.
 [arXiv:2005.02807 [gr-qc]].

\bibitem{wei72}
Weinberg, S.,  {\it Gravitation and Cosmology: Principles and
	applications of the general theory of relativity}, (Wiley, New
York), (1972).

\bibitem{visser05}
M. Visser, 
Gen. Rel. Grav. {\bf 37} , 1541 (2005). 


 \bibitem{Escamilla-Rivera:2019aol} 
  C.~Escamilla-Rivera and S.~Capozziello,
  Int.\ J.\ Mod.\ Phys.\ D {\bf 28}, no. 12, 1950154 (2019)
  doi:10.1142/S0218271819501542
  [arXiv:1905.04602 [gr-qc]].
 
 
 
 \bibitem{wang17} A. Wang, Int. J. Mod. Phys. D {\bf 26}, 1730014 (2017).
 
 \bibitem{cnps09}
 C. Charmousis, G. Niz, A. Padilla and P.M. Saffin, 
 JHEP {\bf 08}, 070 (2009).
 
 \bibitem{bps09}
  D. Blas, O. Pujolas and S. Sibiryakov, 
 JHEP {\bf 10}, 029 (2009). 
 
 
 \bibitem{Sotiriou:2009gy}
 T.~P.~Sotiriou, M.~Visser and S.~Weinfurtner,
 ``Phenomenologically viable Lorentz-violating quantum gravity,''
 Phys.\ Rev.\ Lett.\  {\bf 102}, 251601 (2009)
 doi:10.1103/PhysRevLett.102.251601
 [arXiv:0904.4464 [hep-th]].
 
 \bibitem{Mukohyama:2010xz}
 S.~Mukohyama,
 ``Horava-Lifshitz Cosmology: A Review,''
 Class.\ Quant.\ Grav.\  {\bf 27}, 223101 (2010)
 doi:10.1088/0264-9381/27/22/223101
 [arXiv:1007.5199 [hep-th]].
 
 \bibitem{Izumi:2011eh}
 K.~Izumi and S.~Mukohyama,
 ``Nonlinear superhorizon perturbations in Horava-Lifshitz gravity,''
 Phys.\ Rev.\ D {\bf 84}, 064025 (2011)
 doi:10.1103/PhysRevD.84.064025
 [arXiv:1105.0246 [hep-th]].
 
 \bibitem{Gumrukcuoglu:2011ef}
 A.~E.~Gumrukcuoglu, S.~Mukohyama and A.~Wang,
 ``General relativity limit of Horava-Lifshitz gravity with a scalar field in gradient expansion,''
 Phys.\ Rev.\ D {\bf 85}, 064042 (2012)
 doi:10.1103/PhysRevD.85.064042
 [arXiv:1109.2609 [hep-th]].
 
 \bibitem{Fukushima:2018xgv}
 M.~Fukushima, Y.~Misonoh, S.~Miyashita and S.~Sato,
 ``Stable singularity-free cosmological solutions in nonprojectable Horava-Lifshitz gravity,''
 Phys.\ Rev.\ D {\bf 99}, no. 6, 064004 (2019)
 doi:10.1103/PhysRevD.99.064004
 [arXiv:1812.10295 [gr-qc]].
 
 
 
\bibitem{Ramos:2018oku}
O.~Ramos and E.~Barausse,
Phys. Rev. D \textbf{99} (2019) no.2, 024034
doi:10.1103/PhysRevD.99.024034
[arXiv:1811.07786 [gr-qc]].

\bibitem{Ma:2016lwr}
M.~S.~Ma, R.~Zhao and Y.~S.~Liu,
Class. Quant. Grav. \textbf{34} (2017) no.16, 165009
doi:10.1088/1361-6382/aa8000
[arXiv:1604.06998 [hep-th]].

\bibitem{Gao:2020ddm}
F.~Gao and J.~Llibre,
Eur. Phys. J. C \textbf{80} (2020) no.2, 137
doi:10.1140/epjc/s10052-020-7714-3

\bibitem{Nilsson:2019bxv}
N.~A.~Nilsson,
Eur. Phys. J. Plus \textbf{135} (2020) no.4, 361
doi:10.1140/epjp/s13360-020-00369-w
[arXiv:1910.14414 [gr-qc]].
 
 
 
 
 \bibitem{park09} M.I. Park, J. High Energy Phys. 123, 0909 (2009). 
 
\bibitem{Scolnic:2017caz}
D.~Scolnic, et al
Astrophys. J. \textbf{859} (2018) no.2, 101
doi:10.3847/1538-4357/aab9bb
[arXiv:1710.00845 [astro-ph.CO]].

\bibitem{Jimenez:2019onw}
R.~Jimenez, A.~Cimatti, L.~Verde, M.~Moresco and B.~Wandelt,
JCAP \textbf{03} (2019), 043
doi:10.1088/1475-7516/2019/03/043
[arXiv:1902.07081 [astro-ph.CO]].

\bibitem{Rusu:2019xrq}
C.~E.~Rusu, K.~C.~Wong, V.~Bonvin, D.~Sluse, S.~H.~Suyu, C.~D.~Fassnacht, J.~H.~Chan, S.~Hilbert, M.~W.~Auger, A.~Sonnenfeld, S.~Birrer, F.~Courbin, T.~Treu, G.~C.~F.~Chen, A.~Halkola, L.~V.~Koopmans, P.~J.~Marshall and A.~J.~Shajib,
[arXiv:1905.09338 [astro-ph.CO]].
  
  \bibitem{Kyutoku:2016zxn} 
  K.~Kyutoku and N.~Seto,
  Phys.\ Rev.\ D {\bf 95}, no. 8, 083525 (2017)
  doi:10.1103/PhysRevD.95.083525
  [arXiv:1609.07142 [astro-ph.CO]].
  
  \bibitem{Zhang:2018byx} 
  X.~N.~Zhang, L.~F.~Wang, J.~F.~Zhang and X.~Zhang,
  Phys.\ Rev.\ D {\bf 99}, no. 6, 063510 (2019)
  doi:10.1103/PhysRevD.99.063510
  [arXiv:1804.08379 [astro-ph.CO]].
  
\bibitem{Handley:2019tkm}
W.~Handley,
[arXiv:1908.09139 [astro-ph.CO]].
   
\bibitem{DiValentino:2019qzk}
E.~Di Valentino, A.~Melchiorri and J.~Silk,
Nat. Astron. \textbf{4}, no.2, 196-203 (2019)
doi:10.1038/s41550-019-0906-9
[arXiv:1911.02087 [astro-ph.CO]].

\bibitem{Efstathiou:2020wem}
G.~Efstathiou and S.~Gratton,
Mon. Not. Roy. Astron. Soc. \textbf{496} (2020) no.1, L91-L95
doi:10.1093/mnrasl/slaa093
[arXiv:2002.06892 [astro-ph.CO]].



\end{thebibliography}
\end{document}